\documentclass[pre,twocolumn,superscriptaddress,preprintnumbers,floatfix]{revtex4}

\usepackage{dcolumn}
\usepackage{amsmath}
\usepackage{amssymb}
\usepackage{graphicx}
\usepackage{bm}   % bold math
\usepackage{bbm}   
\usepackage{verbatim}
\usepackage{stmaryrd}
\usepackage{amsthm}

\theoremstyle{break} 	
\theoremstyle{plain} 	
\theoremstyle{break} 	
\theoremstyle{plain} 	
\theoremstyle{break} 	
\theoremstyle{plain} 	
\theoremstyle{plain}	
\theoremstyle{break}	 
\theoremstyle{plain}	
\theoremstyle{break}	 

\def\bra#1{\mathinner{\langle{#1}|}}
\def\ket#1{\mathinner{|{#1}\rangle}}
\def\braket#1{\mathinner{\langle{#1}\rangle}}

\newcommand{\Prob}      {\mathrm{Pr}}

\newcommand{\svector}[1]{\bra{ {#1} }}
\def\Abet {{\mathcal A}}
\def\symb {s}

\begin{document}

\title{Infinite Correlation in Measured Quantum Processes}

\author{Karoline Wiesner}
\email{karoline@cse.ucdavis.edu}
\affiliation{Center for Computational Science \& Engineering and Physics Department,
University of California Davis, One Shields Avenue, Davis, CA 95616}
\author{James P. Crutchfield}
\email{chaos@cse.ucdavis.edu}
\affiliation{Center for Computational Science \& Engineering and Physics Department,
University of California Davis, One Shields Avenue, Davis, CA 95616}

\bibliographystyle{unsrt}

\begin{abstract}
We show that quantum dynamical systems can exhibit infinite correlations
in their behavior when repeatedly measured. We model quantum processes
using quantum finite-state generators and take the stochastic language
they generate as a representation of their behavior. We analyze two
spin-$1$ quantum systems that differ only in how they are observed. The
corresponding language generated has short-range correlation in one
case and infinite correlation in the other.
\end{abstract}

\maketitle

We study how sequences produced by a \emph{quantum information source}
can produce infinite-length correlations.
To start, we recall the finitary \emph{quantum generators}
defined in Ref. \cite{wies:06b}. They consist of a finite set of
\emph{internal states} $Q = \{q_i: i = 1, \ldots, |Q| \}$. The
\emph{state vector} is an element of a $|Q|$-dimensional Hilbert
space: $\bra{\psi} \in \mathcal{H}$. At each time step a
quantum generator outputs a symbol $s \in \Abet$ and updates its
state vector as follows.

The temporal dynamics is governed by a set of $|Q|$-dimensional
\emph{transition matrices} $\{T(s) = U \cdot P(s), s \in \Abet \}$,
whose components are elements of the complex unit disk and where each
is a product of a unitary matrix $U$ and a projection operator $P(s)$.
$U$ is a $|Q|$-dimensional unitary \emph{evolution operator} that
governs the evolution of the state vector. $\mathbf{P} =\{ P(s): s \in \Abet \}$
is a set of \emph{projection operators}---$|Q|$-dimensional Hermitian
matrices---that determines how the state vector is measured.

The output symbol $s$ is identified with the measurement outcome and
labels the system's eigenstates. The projection operators determine how output symbols are generated
from the internal, hidden unitary dynamics. They are the only way to observe
a quantum process's current internal state. 

We can now describe a quantum generator's operation. $U_{ij}$ gives the
transition amplitude from internal state $q_i$ to internal state $q_j$.
Starting in state vector 
$\bra{\psi_0}$ the generator updates its state by applying the
unitary matrix $U$. Then the state vector is projected using $P(s)$ and
renormalized. Finally, symbol $s \in \Abet$ is emitted. In other words,
starting with state vector $\svector{\psi_0}$, a single time-step yields
$\bra{ \psi(s) } = \bra{\psi_0} U \cdot P(s)$.

An observer is interested in what can be observed and these are the
measurement outcomes $s \in \Abet$. Thus, the only way to describe a
quantum process is in terms of the sequence
$\stackrel{\leftrightarrow}{S} \, \equiv \,\ldots S_{-2} S_{-1}
S_0 S_1 \ldots$
of observed random variables $S_t$ produced by a quantum generator. We
consider a family of distributions, 
$\{ {\rm Pr}(s_{t+1} , \ldots , s_{t+L}): s_t \in \Abet \}$, where
${\rm Pr}(s_t)$ denotes the probability that at time $t$ the random
variable $S_t$ takes on the particular value $s_t \in \Abet$ and
${\rm Pr} (s_{t+1} , \ldots , s_{t+L})$ denotes the joint probability over
sequences of $L$ consecutive measurement outcomes. We assume that the
distribution is stationary;
${\rm Pr}(S_{t+1},\ldots, S_{t+L})={\rm Pr}(S_1, \ldots , S_L )$. 
We denote a block of $L$ consecutive variables by $S^L \equiv S_1 \ldots S_L$
and the lowercase $s^L = s_1 s_2 \cdots s_{L}$ denotes a particular
measurement sequence of length $L$. We use the term {\em quantum process}
to refer to the joint distribution
${\rm Pr} (\stackrel{\leftrightarrow}{S})$ over the infinite chain of
random variables. A quantum process, defined in this way, is the quantum
analog of what Shannon referred to as an \emph{information source} \cite{cover}. 
We can now determine word probabilities of observations of a
\emph{quantum finite-state generator} (QFG). Starting the generator
in $\bra{\psi_0}$ the probability
of output symbol $s$ is given by the state vector without
renormalization:
\begin{equation}
\label{eqn:qpry}
\Prob(s) =  \braket{ \psi(s) | \psi(s)} ~.
\end{equation}
The probability of outcomes $s^L$ from a measurement sequence is
\begin{equation}
\label{eqn:qprsL}
\Prob(s^L) =  \braket{ \psi(s^L) | \psi(s^L)} ~.
\end{equation}

We will now investigate word probabilities of a particular quantum process.
Consider a spin-$1$ particle that is subject to a magnetic field
which rotates the spin. The state evolution can be described by the
following unitary matrix:
\begin{equation}
U = \left(
	\begin{array}{ccc}
		\frac{1}{\sqrt{2}} & \frac{1}{\sqrt{2}} & 0
		\\ 0 & 0 & -1
		\\ -\frac{1}{\sqrt{2}} & \frac{1}{\sqrt{2}} & 0 
	\end{array}
	\right) ~.\\
\label{eqn:qgldm}
\end{equation}
Geometrically, $U$ defines a rotation in $R^3$ around the y-axis by
angle $\frac{\pi}{4}$ followed by a rotation around the x-axis by
an angle $\frac{\pi}{2}$.

Using a suitable representation of the spin operators $J_i$ \cite{messiah}
such as:
\begin{align}
J_x & = \left(
    \begin{array}{ccc}
        0 & 0 & 0 \\
        0 & 0 & i \\
        0 & -i & 0
    \end{array}
    \right) ,~
J_y = \left(
    \begin{array}{ccc}
        0 & 0 & i \\
        0 & 0 & 0 \\
        -i & 0 & 0
    \end{array}
    \right) ,~ \nonumber \\
J_z & = \left(
    \begin{array}{ccc}
        0 & i & 0 \\
        -i & 0 & 0 \\
        0 & 0 & 0
    \end{array}
    \right) ,
\end{align}
the relation $P_i = 1 - J_i^2$
defines a one-to-one correspondence between the projector $P_i$ and the
square of the spin component along the i-axis. The resulting measurement
represents the yes-no question, Is the square of the spin component along
the $i$-axis zero?

Consider the observable $J_y^2$. Then the following projection operators
define the quantum finite-state generator:
\begin{equation}
P(0) = \ket{010}\bra{010}
 ~\mathrm{and}~
~P(1) = \ket{101}\bra{101}
 ~.
\end{equation}

The stochastic language generated by this process is the so-called
\emph{Golden-Mean Process} language \cite{crutchfield:03}. It is
defined by the set of \emph{irreducible forbidden words}
$\mathcal{F} = \{00\}$. That is, no consecutive zeros occur. For
the spin-$1$ particle this means that the spin component along the
y-axis never equals 0 twice in a row. We call this
\emph{short-range correlation} since there is a correlation between a
measurement outcome at time $t$ and the immediately preceding one
at time $t-1$. If the outcome is $0$ the next outcome will be
$1$ with certainty. If the outcome is $1$ the next measurement is
maximally uncertain: outcomes $0$ and $1$ occur with equal probability.

\begin{figure}
\begin{center}
\resizebox{1.80in}{!}{\includegraphics{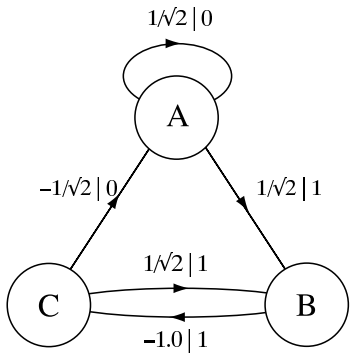}}
\end{center}
\caption{The Even Process quantum generator.
  }
\label{fig:ep-qdg}
\end{figure}

\begin{figure}
\begin{center}
\resizebox{!}{2.00in}{\includegraphics{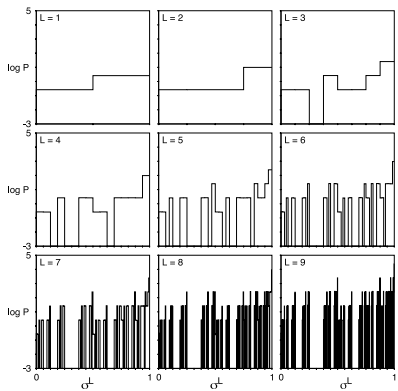}}
\end{center}
\caption{Even Process stochastic language: Words $\{01^{2k-1}0\}, k=1,2,3,...$ 
  have zero probability; all others have nonzero probability. The logarithm
  base 2 of the word probabilities is plotted
  versus the binary string $\symb^L$, represented as base-$2$ real number
  ``$0.\symb^L$''. To allow word probabilities to be compared at different
  lengths, the distribution is normalized on $[0,1]$---that is, the
  probabilities are calculated as densities.
  }
\label{fig:ep-lang}
\end{figure}  

Consider the same Hamiltonian, but now use instead the observable $J_x^2$.
The corresponding projection operators define the QFG:
\begin{equation}
P(0) = \ket{100}\bra{100}
 ~\mathrm{and}~
~P(1) = \ket{011}\bra{011}
 ~.
\end{equation}
The QFG defined by $U$ and the above projection operators is shown in
Fig.~\ref{fig:ep-qdg}.
The stochastic language generated by this process is the so-called
\emph{Even Process} language \cite{crutchfield:03}. The word distribution
is shown in Fig.~\ref{fig:ep-lang}. It is defined by the infinite set
of irreducible forbidden words $\mathcal{F} = \{01^{2k-1}0\}, k=1,2,3,...$.
That is, if the spin component equals 0 along the y-axis it will be zero
an even number of consecutive measurements before being observed to be
nonzero. This is where the infinite correlation is found: For a possibly
infinite number of time steps the system tracks the evenness or oddness
of number of consecutive measurements of ``spin component equals 0 along
the y-axis''.

% **********************************************************************
%\section{Final Remarks}

The above examples show that quantum dynamical systems store information
in their \emph{behavior}. The quantum Even Process example is particularly
striking since it has only three internal states, but exhibits infinitely
long temporal correlations. Comparing the two examples demonstrates, in
addition, that the amount of stored information depends on the means taken
to observe the system. These properties are quantified by adapting
information-theoretic measures of randomness and memory to quantum processes.
We demonstrated that a repeatedly measured quantum dynamical system can
store, in its current state, information about previous measurement outcomes.

In quantum computation the experimentalist subjects information stored in
an initially coherent set of physical degrees of freedom to a selected
sequence of manipulations. The system's resulting state is measured and
interpreted as the output of a computation. Our computation-theoretic
approach, in contrast, applies to continuous computation and shows how
information processing is embedded in even simple quantum systems' behavior.
\vspace{-0.3in}

% **************************** REFERENCES ****************************
\bibliography{ref}

\begin{thebibliography}{1}

\bibitem{wies:06b}
K.~Wiesner and J.~P. Crutchfield.
\newblock Computation in finitary quantum processes.
\newblock 2006.
\newblock e-print arxiv/quant-ph/0608206.

\bibitem{cover}
T.~Cover and J.~Thomas.
\newblock {\em Elements of Information Theory}.
\newblock Wiley-Interscience, 1991.

\bibitem{messiah}
Albert Messiah.
\newblock {\em Quantum Mechanics}, volume~II.
\newblock John Wiley \& Sons, Inc. -- New York, 1962.

\bibitem{crutchfield:03}
J.~P. Crutchfield and D.~P. Feldman.
\newblock Regularities unseen, randomness observed: Levels of entropy
  convergence.
\newblock {\em Chaos}, 13:25 -- 54, 2003.

\bibitem{wiesner:06}
K.~Wiesner and J.~P. Crutchfield.
\newblock Language diversity in measured quantum processes.
\newblock {\em Intl. J. Unconventional Computation}, 2006.
\newblock in press.

\end{thebibliography}
\end{document}